\documentclass[letterpaper,10pt]{article}

\usepackage{stix}
\usepackage{setspace}
\usepackage[small]{titlesec}
\usepackage{xspace}

\usepackage[utf8]{inputenc}
\usepackage{graphicx}
\usepackage[export]{adjustbox}
\usepackage{placeins}
\usepackage{url}
\usepackage{listings}
\usepackage{enumerate}
\usepackage{doi}
\usepackage[numbers,sort&compress]{natbib}
\usepackage{subcaption}
\bibpunct{\nolinebreak{}[}{]}{,}{n}{}{,}
\usepackage[margin=1in]{geometry}
\usepackage{float}
\usepackage{xspace}
\usepackage{booktabs,multirow,array,makecell}
\usepackage{caption}

\renewcommand{\figurename}{Fig.}

\usepackage[scaled=0.96]{zi4}
\title{An Ontological Metamodel for Cyber-Physical System Safety, Security,
       and Resilience Coengineering}

\date{}
\renewcommand\footnotemark{}
\author{Georgios Bakirtzis \and
        Tim Sherburne \and 
        Stephen Adams \and
        Barry M. Horowitz \and
        Peter A. Beling \and
        Cody H. Fleming
        \thanks{G. Bakirtzis, T. Sherburne, S. Adams, B.M. Horowitz, P.A. Beling, and C.H. Fleming are with the University of Virginia. Corresponding author: \texttt{bakirtzis@virginia.edu}}
}

\begin{document}

\maketitle

\begin{abstract}
System complexity has become ubiquitous 
in the design, assessment, and implementation of practical
and useful cyber-physical systems.
This increased complexity is impacting the management
of models necessary for designing cyber-physical systems
that are able to take into account a number of ``-ilities'',
such that they are safe and secure and ultimately resilient to disruption of service.
We propose an ontological metamodel for system design
that augments an already existing industry metamodel to capture the relationships between various model elements and safety, security, and resilient considerations.
Employing this metamodel leads to more cohesive
and structured modeling efforts
with an overall increase in scalability, usability,
and unification of already existing models.
In turn, this leads to a mission-oriented perspective
in designing security defenses and resilience mechanisms
to combat undesirable behaviors.
We illustrate this metamodel in an open-source GraphQL implementation,
which can interface with a number of modeling languages.
We support our proposed metamodel with a detailed demonstration
using an oil and gas pipeline model.
\end{abstract}

\section{Introduction}
% discuss mission-oriented secure & the need for modeling with the metamodel

% space & weapon systems compliance - ?? stpa standard

% breadth and depth of model? cost / scope trade-off
% hierarchical / iterative 

% model validation for completeness check

% models as living version controlled document

% tails of a distribution (BH)

% solutions are narrower the problems are broader
Cyber-physical systems (CPS) integrate diverse software, firmware, and hardware
to control the operation of the physical components of the system based on analysis of sensor data.
These largely distributed systems have already
sparked innovation in a number of sectors, 
such as agriculture, power distribution,
search and rescue, and weaponry.
In the face of rapid innovation, it is important
to understand and address the new challenges
that arise from the deployment of CPS in environments where they are subject to cyber attacks.
In particular, a characteristic of CPS is that when they fail, either because of an intrinsic fault
or a security violation, they can transition to a hazardous state
and, therefore, can cause accidents or otherwise unacceptable losses.
In the domain of CPS, therefore, cybersecurity and safety are intertwined~\cite{penzenstadler:2014}.
However, safety and security are often viewed disjointly in a way that considers security mitigations without regard to safety losses~\cite{bakirtzis:2020}. The classical approach to security involves verification of components and the construction of security barriers. The most effective approaches to safety, by contrast, center on modeling losses in terms of control actions with respect to states that may reflect the condition of many components \cite{leveson2018stpa}.  We aim to combine elements from the security and safety parardims to develop models to support the design of resilience solutions that can mitigate against unacceptable losses.
Resilience is a proactive notion where in response
to anomalies the system has a way to maintain operation, even in some degraded form \cite{hosseini:2016}.
The application of resilience concepts to safety and security design has the potential
to transform our safety considerations
and security defenses from a prescriptive, perimeter-based approach
to one that uses adaptable and dynamic design patterns that take into account the intended functions of the system. Such a transformation can be achieved only if we are able to unify safety, security, and resilience models.

The intrinsic complexity of CPS creates difficulties regarding how to best account for and model the intertwined nature of safety, security, and resilience \cite{lee:2016}.
Modeling safety requires thinking in terms of misbehaviors, unacceptable losses, and hazardous states.
Modeling security requires thinking in terms of confidentiality, integrity, and availability.
Modeling resilience requires thinking in terms of redundant architectures, recovery from a death state, and restricting the expected service of the system to disallow possible violations.
Safety, security, and resilience have significant overlap in CPS applications.
For example, both a security violation and a safety related component failure could cause an unsafe operator control action, resulting in a  system response that reduces the dependability of the overall system.
While this example of CPS behavior clearly shows that security and safety are coupled, the corresponding model representations are often decoupled and disparate.
To combat this disjoint treatment of safety, security, and resilience, modeling efforts must be structured in terms that not only address all three qualities but also are able to relate their results to each other.

Structured modeling efforts are often captured in a metamodel,
which defines the \emph{types} and \emph{relationships}
of information allowed in a model of a system.
Different metamodels apply to different applications, for example,
a metamodel for geologic maps is not going to be the same
as the metamodel for a software application.
A strict metamodel would enforce the types and relationships
but often metamodels are used to provide a structured guidance for model creation.
If automated, such a metamodel can be used
to assist systems designers the same way that a sophisticated text editor does for programmers.
In practice, this capability is still lacking in modeling tools,
predominantly because metamodels are not developed for a particular application
and, therefore, cannot enforce strict rules to users
that have a diverse set of metamodel requirements.
The ontological study of metamodels augments them
with a refinement of types.
Ontologies define what entities might exist
in a particular domain application, how they might be grouped
in categories, and how they relate but also can be decomposed within a hierarchy of types.
The metamodel, then, is a representation of this ontology, often represented in a set of graphs.

Ontologies and metamodels can be particularly useful
in the design of safety-critical CPS.
For any design solution, the multitude of diverse but equally important system representations requires management of their corresponding models and relationships among them. This is challenging when attempting to manage some of the most concerning system quality attributes, such as safety and security, as early as possible in the systems lifecycle. Usage of models for this purpose can be approached in a number of ways to transition from assuring safety, security, and resilience by trial and error to addressing them early through the use of models. While other ``-ilities''~\cite{voas:2004}, for example dependability, survivability, and reliability have been addressed through models, safety, security and resilience have not been addressed as one problem. However, in CPS these three qualities are perhaps the most important for the eventual deployment and use of these systems.

To achieve this transition it requires us to learn (but avoid repeating) from highly resilient systems built over time, such as those developed with a fly--fix--fly approach in the United States Federal Aviation Administration's air traffic control system. In that case, the system modifies aircraft flight rules depending on pilot visibility through use of instrument flight rules and visual flight rules, thereby, as necessary, reducing airport capacity while increasing safety. But this raises the question about how should systems engineers model ``-ilities'' of the system, which are interconnected and cannot be examined in isolation, before having experienced them during deployment? As a specific example, how would we model the interconnection of a safety incident that was initiated by a security violation in the case of CPS? And further, how would one design resilience and relate such resilience solutions to losses and their prevention to assure the proper operation of complex CPS systems during deployment?

Solutions to parts of these questions have been proposed
in each individual field of safety, security, and resilience
but without recognizing the problems that arise
when these metrics interact with each other.
For example, in the field of safety, Leveson has developed systems-theoretic accident model and process (STAMP)~\cite{leveson:2011},
which examines safety incidents in terms
of hierarchical control and unacceptable losses.
In the intersection of security and resilience there is System Aware~\cite{jones:2012,jones2013architectural}
and its evolution Mission Aware~\cite{carter:2018,carter:2019}, which see mitigation
in terms of resilient modes in addition
to traditional security defenses.
Finally, the general field of model-based systems engineering (MBSE)
has produced both methods and tools in assessing the safety
and security posture of system designs
and is the paradigm in which the majority
of system models must conform to.
An example of a fundamental systems engineering metamodel is the model
openly published by Vitech Corporation in support of the company's modeling software~\cite{vitech2018onemodel}.

One unifying answer is to create pragmatic ontologies based 
on the above capabilities developed
within each individual field,
which will allow for a holistic examination
of those qualities in relation to the system model.
Crucially, in this paper we extend an already existing and often used
systems engineering metamodel
to address safety, security, and resilience.
For modeling purposes, this ontology is often codified in metamodels.
Previous works study the theory of ontologies extensively
for the purpose of understanding them
at a meta-meta level \cite{obrst:2003, karagiannis:2006, onggo:2010, morozov:2018}.
However, there is a gap 
in creating domain specific ontologies
and metamodels; especially for the modeling of CPS.
Creating an ontology in the field of CPS should take
into account both the system and its associated ``-ilities''
with a particular focus on safety, security, and resilience.
Assurance of these three ``-ilities'' combats undesirable behaviors
during deployment by focusing the design effort
both on the functional and non-functional requirements
of the system under design.
Therefore, a concrete implementation of an ontological metamodel
is one way forward in multi-paradigm modeling for CPS,
which is a necessity to manage the intrinsic complexity
of these systems \cite{vangheluwe:2019}.

For these reasons, we propose the use of an ontological metamodel that can be applied to the differing modeling views necessary to design CPS. We assert that this metamodel supports the structuring of modeling efforts. Additionally, this metamodel can help users see interoperability between various models and assessment methodologies, which has the potential of documenting both formal and informal requirements in one model \cite{golra:2018} and ultimately achieve model federation. There are significant design risks if we permit reliability, safety, and security engineers to work in isolation when designing highly automated CPS. Instead, such  attributes must be examined together throughout lifecycle such that appropriate tradeoffs and design decisions can also be made throughout the lifecycle. The metamodel can act as a bridge between system models and safety, security, and resilience by providing a common language that is understood by all stakeholders of the system. Thus a metamodel can be used to support and foster collaborative design.

We recognize that a single ontology that addresses system issues associated with a large number of quality attributes would be cumbersome and likely not practical. Therefore, we do not seek to develop  a universal ontology that would be applicable in all applications; different ontologies are often necessary as they provide differing views. However, we are developing our suggested ontology that integrates safety, security, and resilience considerations on top of an already tested metamodel, for the particular purpose of providing a mission-oriented perspective to system modeling and assurance.

Our suggested ontology will be one that addresses system engineering  for systems where safety, security, and resilience are considered to be critical  quality attributes. To make our contributions concrete in this area we have implemented a GraphQL~\cite{graphql} definition of this ontological metamodel in the form of a generalized schema. GraphQL is a vendor agnostic format for schema specification augmented with a query language. Therefore, GraphQL implements the ontological metamodel as a schema that can be used by any modeling language, given minimal export capability.

In summary, our contributions are as follows:
\begin{itemize}
 \item We derive a metamodel for structured system modeling that addresses safety and security and resilience as one problem. Specifically, we build on and adapt STAMP and mission aware cybersecurity to create general connections between safety and security concepts.
\item We use the metamodel to assist with a tradespace analysis of resilience solutions to the losses associated with safety and security violations.
\item We implement this metamodel in GraphQL, which encodes these relationships and attributes formally and can interface with a number of tools.

% from conclusions
% By implementing this unification in a concrete metamodel we facilitate consistency between CPS model views; coordination of safety, security, and resilience with system models; and tradespace analysis of these three metrics in relation to candidate design solutions.
% 

\end{itemize}

By achieving this unification we are able to promote consistency between model views, provide one possible solution to the coordination of safety, security, and resilience within a system model, and are able to apply tradespace analysis of design solution with regard to these non-functional requirements. We illustrate the use of the metamodel in a oil and gas pipeline demonstration. 

% -ilities have a lot of overlap. We are going to start by looking at safety and security. Factors for relating to the overlap.

% Specialization in ilities but, for example, a security violation can affect any of the ilities.

% Metamodel tells you what the model has to be.
\section{Background}
The proposed metamodel is an integration of multiple technology areas
into a unified ontology. 
This integration leverages advances in MBSE \cite{leibrandt:2001}, safety through STAMP \cite{leveson:2011}, and mission aware cybersecurity \cite{carter:2018}.
Mission aware is a systems engineering methodology that attempts
to bridge system design with safety, security,
and resilience.
We consider the mission aware metamodel a specific application
of the metamodel and not necessarily defining the use
of the metamodel itself.
Our general motivation is to construct a bridge between modeling paradigms and verification tools, which stems from the need to build tool interoperability \cite{broy:2010}.
For example, adding a contract-based framework~\cite{dragomir:2017}
to already existing tools is largely a manual process, which could be assisted
by enforcing a structured modeling framework through metamodeling.
Generally, by using a consistent metamodel such model transformations become part of the system design effort, which is an increasingly desirable quality~\cite{lucio:2014}.
This observation is consistent with surveys of the modeling community:
consistent and automated model transformations that are language agnostic
are necessary to transition model based design in industrial practice~\cite{bruel:2020}.
The algorithmic implementation of the metamodel provides one answer to modeling tool interoperability, which has the potential to create connections between static views
of the system, for example, SysML block diagrams, and dynamics views of the system, such as Mathworks Simulink.
In turn, this could assist with providing evidence for the safe operation of CPS, which requires the use of several types of models \cite{mitra:2013}.

\subsection{Model-Based Systems Engineering}

MBSE has gathered increasing attention
both by the United States Department of Defense~\cite{dod:2018}
and by standards committees, for example, 
RTCA DO-331~\cite{DO-331} and RTCA DO-333~\cite{DO-333},
and SAE AS5506C~\cite{AS5506C}, to name a few.
The tenant of MBSE is that models take a central role
in the design and development of systems.
Even further, models take the role of \emph{living documents}, where justifications
for design choices and changes are recorded and disseminated.
In addition, MBSE attempts to bridge potential system design gaps 
by encapsulating the several layers of abstraction necessary
in the design of systems.
The assertion is that different views of a system's design must be contrasted and related.
For example, candidate implementations of the system are modeled
in relation to desired behaviors and requirements.
The promise of MBSE is that within a modeling language one can diagrammatically
represent any given hierarchical level of system design, from requirements
to functional behaviors to architectures.

One such language is systems modeling language (SysML). SysML is a general-purpose specification of a modeling language for systems 
engineering which has been maintained by the object management group (OMG) since 2007 \cite{sysml}.
Different vendors have implemented the SysML standard and, therefore, there is an abundance
of tools available in the market.
As a standard, SysML allows for a variety of systems engineering approaches
to be developed within what might be considered a metamodel specification
for system design \cite{wolny:2020}.
In 2017 the OMG issued a request for proposals for a second generation specification known as SysMLv2 \cite{sysmlv2}. 
A primary requirement for SysMLv2 is to augment the flexible syntax of block diagrams 
with an overall model representing best systems engineering practice.
This demonstrates the need for a concrete metamodel for system design that is used to control relationships between model entities and views.

Academia has already provided a number of solutions to this need \cite{kalnins:2019,atkinson:2002,atkinson:2015}.
However, academic solutions that attempt to view the system holistically often lack formal semantics, without which the solution cannot provide the necessary consistency between model views.
On the other hand, solutions implementing formal semantics often address a small subset of the overall system design.
In this paper, we attempt to build a metamodel that both views the system holistically
and implements formal semantics for all fundamental entities and views of system design,
including entities relating to safety, security, and resilience.

To achieve this merger, we build upon a metamodel
that is already successfully used in industry.
Specifically, we use the Vitech metamodel \cite{vitech2018onemodel} as a starting point.
The Vitech metamodel is publicly available and is well aligned with the SysMLv2 requirement for ``precise systems engineering semantics.''
Our augmentations to this industry metamodel include a number of entities and interactions
to address safety, security, and resilience, which are necessary for the design and deployment of CPS.
This stemmed from the challenge of extending beyond the system itself
and being able to capture mission-oriented information.
It is possible to bring ``front and center'' a number of whys, hows, and whats
that otherwise would be missed by just examining the system
in isolation of its purpose.
Often a system designer's ``whats'' might be a software engineers how's \cite{whalen:2012}.
A metamodel explicitly considers the hierarchy within system design,
where at each layer there is both behavior and architecture,
an important distinction compared to modeling languages
that view the two as distinct.

\begin{figure*}[h]
\centering
\includegraphics[width=\linewidth] {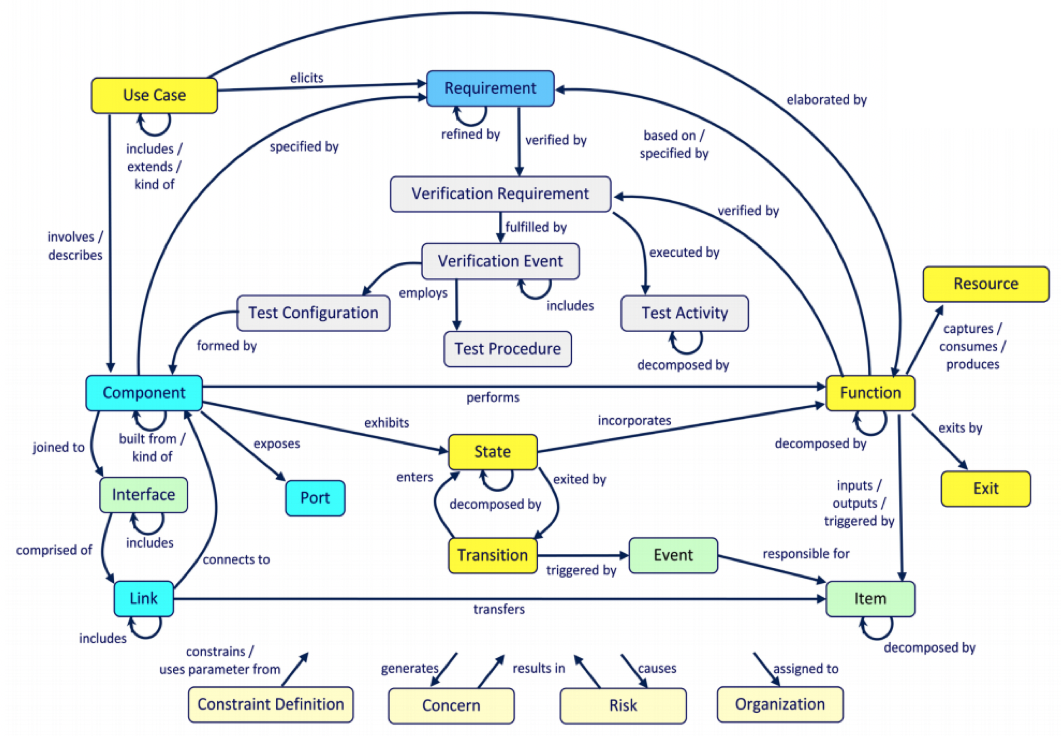}
\caption{The Vitech model-based systems engineering metamodel is suited to build upon a general metamodel for cyber-physical system design, safety, security, and resilience  (reproduced from Scott and Long~\cite{vitech2018onemodel}).}
\label{fig:vitech-mbse}
\end{figure*}

Specifically, what we consider to be the MBSE metamodel represents critical systems engineering concepts and their interrelationships spanning requirements, behavior, architecture, and testing (Fig.~\ref{fig:vitech-mbse}).  This integrated model presents a high-level view of not only the ultimate specification of a system, but also the journey to that specification -- concerns opened and closed, risks identified and managed. As a mechanism for managing system complexity, the metamodel supports an incremental, layered approach via the consistent relationship across requirements, behavior, architecture, and test domains.

The domain of CPS is particularly suited for exercising MBSE
because they rely on both analogue behavior -- the physics --
and digital -- the computation.
Often modeling tools are disjoint and they allow the designer to either allow the designer to, for example, simulate the physics of the system on one hand or alternatively, verify correctness of software on the other hand.
Because CPS brings both physics and computation together in one system, 
it requires use of models that address both physical and computational performance as a coupled system.
This consists of one of the main challenges in modeling CPS,
namely the challenge to capture all interactions between those two different domains.
A solution to this challenge is to enforce clear semantics between model entities,
such that high-fidelity modeling is possible.
This is the predominant motivation for creating a language-agnostic metamodel
that contains relationships between standard modeling paradigms, for example,
behavioral and architectural entities, and necessary performance metrics
for assuring the correct behavior of CPS, for example, addressing both safety and security concerns.

\subsection{Loss Identification with STAMP} \label{Section: STPA}

Safety assessment has a long tradition.
Here we present some of the results coming from STAMP,
a safety analysis method that is based on causation.
Causation in STAMP is modeled through hierarchical control,
which models each level of a system as a controls process,
where unsafe control actions can occur.
This layered approach to safety has the advantage
that unsafe control actions at each level percolate upwards
or downwards in the hierarchy
that in turn provides a notion of consequence
within the safety model.
STAMP works in contrast to linear failure modes,
where unsafe actions form a chain of events.
Instead, in STAMP safety violation emerge
from the interacting control layers governing the system.

Specifically, STAMP \cite{leveson2018stpa} is a hazard analysis technique based
on an extended model of accident causation.
In addition to component failures, STAMP assumes
that accidents can also be caused by unsafe interactions
of system components, none of which may have individually failed.
For this reason, STAMP further asserts that emergent properties, for example safety and security, cannot be assured
by examining subsystems in isolation.
STPA (System-Theoretic Process Analysis) is one flavor of STAMP modeling that is primarily used
to proactively identify hazardous conditions and states.
STPA-Sec is an extension on STPA with the intention of transitioning the benefits of loss-oriented safety assessment to security~\cite{young:2013,young:2014}.

The hierarchical control notion within STAMP
is a congruent idea with a number of MBSE block diagrams 
such as architectural or behavioral diagrams
because they can be augmented to model unsafe control actions
in addition to the control system that define the behavior
and architecture of the system.
Furthermore, MBSE is based on the same hierarchical notion,
namely, that systems can be modeled through different views
that reside in different levels of abstraction.
STAMP entities are, therefore, an important view
of safety and security, which we would like to introduce
in more general form within the metamodel.

Another reason to use notions from STAMP in a metamodel for CPS is that it has recently
been incorporated as a possible safety and security analysis technique
in several standards, for example, SAE J3187 \cite{SAE:J3187}, SOTIF ISO/PAS 21448 \cite{PAS:21448},  RTCA DO-356A \cite{DO:356}, ASTM WK60748 \cite{ASTM:WK60748}, and SAE AIR6913 \cite{SAE:AIR6913}, to name a few.
By grounding the safety and security metamodel entities in existing and future standards,
we encourage a modeling methodology that can assist
with the eventual system certification and validation process.
While we focus our demonstration on STAMP, the construction
of the safety entities within the metamodel are general
and therefore can assist or otherwise augment other hazards or safety methods.

% \begin{figure}[ht]
% \centering
% \includegraphics[width=\linewidth]{Figures/stpa-overview.png}
% \caption{STPA Process Overview (reproduced from Thomas and Leveson \cite{leveson2018stpa}).}
% \label{fig:stpa-overview}
% \end{figure}

% \begin{figure}[ht]
% \centering
% \includegraphics[scale=.4]{Figures/stpa-trace.png}
% \caption{STPA Entity Traceability (reproduced from Thomas and Leveson \cite{leveson2018stpa}).}
% \label{fig:stpa-trace}
% \end{figure}

\subsection{Mission Aware Cybersecurity} \label{section-mission-aware}

\begin{figure*}[th]
\centering
\includegraphics[width=\linewidth]{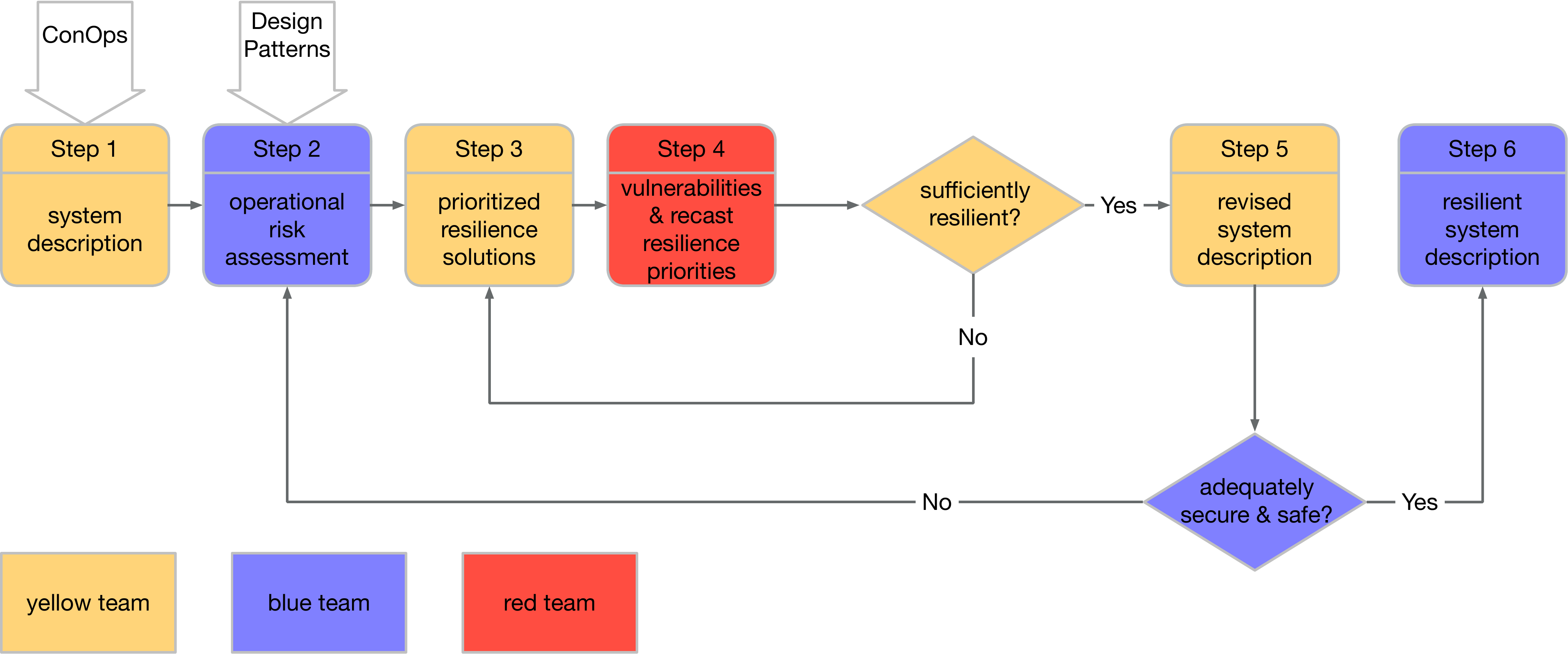}
\caption{In mission aware the systems engineering lifecycle is segmented by specific teams that address system design and resilience based on losses related to safety and security  (adapted from Carter et al. \cite{carter:2019}).}
\label{fig:csrm}
\end{figure*}

Although the metamodel presented in this paper can be used generally for CPS safety, security, and resilience, it is important to provide context for {\em how} the metamodel can be used. One such methodology, termed mission aware, is presented to motivate one example of the metamodel's use (Fig.~\ref{fig:csrm}).
In general, mission aware is a series of steps and processes that brings together expertise from three disparate disciplines: operations (blue team), systems engineering (yellow team), and cybersecurity (red team). The blue team consists of operationally-oriented experts that are familiar with similar current systems and/or handling systems under duress from environmental disturbances, cyber attacks, or other disruptions.The systems engineering team consists of system analysts and modelers with technical expertise. This team is responsible for the development of initial system designs, consequence analysis based on loss scenarios, and the derivation of potential resilience design features. The red team consists of expertise in both defensive and offensive cyber capabilities. 

Previous work \cite{carter:2019} identified a six step approach for engineering in safety, security, and resilience could take the following interaction between model artifacts and teams. 

\begin{enumerate}[Step 1)]
\item	Generate system description produced by the systems engineering team, including objectives and goals, high-level technical requirements, system architecture, and functional or behavioral description in a tool-based medium such as SysML.
\item	Using those model artifacts the blue team conducts a consequence analysis, which results in a prioritized list of undesirable functional outcomes from an operational perspective, as well as a systems engineering team consequence analysis (for example by using STPA).
\item The systems engineering team derives potential resilience solutions based on the results of the previous step.
\item The red team identifies loss scenarios for the system and prioritizes defenses, resilience, and software engineering solutions.
\item The systems engineering team refactors the system descriptions based on red team recommendations.
\item Finally, the blue team responds to the refactored system descriptions.
\end{enumerate}

Resilience in mission aware defines the type of mitigation patterns
that allow the system to maintain state awareness,
which further requires to \emph{proactively} maintain a safe level
of operational normalcy in response to anomalies, including attacks~\cite{rieger:2009}.
A concept familiar to resilient design pattern practitioners
and necessary to achieve this operational normalcy is the \emph{monitor}~\cite{horowitz:2020}.
In mission aware, the monitor is termed a sentinel.
The sentinel's purpose is to be as simple of a system as possible
with provable guarantees on its safe and secure behavior.
In terms of the overall system, the sentinel monitors the state
of the system and given a number of resilient design patterns decides
on mitigative actions if things are measured to be abnormal -- given an initial notion
of what normal behavior means in the design phase and is updated throughout the rest
of the lifecycle.
In the rest of the paper we will use sentinel
to indicate this design pattern.

There are many other possible methodologies and uses of the metamodel, but this section serves as a relatively concrete guide for its use. The brief familiarity with mision aware is necessary for the case study demonstration of the metamodel (Section~\ref{sec:demo}), because the modeling artifacts assume this engineering methodology.

\subsection{Metamodels for System Design}

The problem of creating useful metamodels for software system design
and the overall need for \emph{systems}-level metamodels is not new, 
and indeed the long-term vision of using metamodels to integrate methods and tools
and maintain consistency within different model views has not changed \cite{poole:2001}.
Beyond the Vitech metamodel, which was built specifically
to address an industry need, there are other approaches
to structured modeling that have emerged in the past two decades.
The consensus and general trajectory of this line of research seems
to be towards specialized ontologies for a particular application
and not a general metamodel for all system design.
Specialized ontologies can bridge diverse views necessary
for the particular type of system, in this case CPS, leading
to better management of already existing model types.
In particular, by specializing, one can capture precise semantics for hierarchical decomposition,
which we posit is one way to achieve multilevel modeling \cite{lara:2014}.
Any graphical language can be described
or otherwise contained by the OMG meta object facility (MOF) specification \cite{MOF}.
However, MOF relates to the implementation
of diagrammatic languages and not to the structured modeling we want to facilitate
to assure a number of ``-ilities''.
With this in mind, we see a gap in structured modeling
for the domain of CPS, where we need to provide evidence
of safe and secure operation before deployment.

Practical reports on modeling methods show that currently
we lack effective collaboration, are error-prone in syncing models,
and often metamodels are only useful if used fully \cite{gomez:2020}.
Our metamodel partially overcomes these challenges by working across design artifacts,
allowing checks of agreement by using a consistent metamodel outside
of a particular modeling language or tool, and
does not require full conformance to assist in decision-making.
This means that depending on the CPS application, system modelers can use the appropriate subgraph
of the overall metamodel.

In general, expertly constructing metamodels requires an iterative process \cite{cho:2011}.
Informal modeling tools must be connected to concrete semantics.
These semantics often come from the relationships between metamodel entities
and not necessarily from the entities themselves.
Beyond the actual construction of a metamodel there is  still no deep understanding
about what information a metamodel should include, how it should be included,
and in what way it should be related \cite{williams:2013}.
This means that a large part of metamodeling construction happens by example.

A particular problem requires a particular \emph{type} of entity
and that entity is added to the metamodel.
This can end up creating massive metamodels
that are difficult to navigate and apply during modeling.
This notion of type has led to a number
of works in metamodeling that apply formal semantics
from programming languages, which attempt to manage this process \cite{berg:2012,berg:2014,berg:2015}.
Composition of type entities can then lead
to new types without extending the base metamodel.
Composition of different metamodels can lead
to modeling tool interoperability \cite{combemale:2010},
which speaks to the need to create small compact metamodels
for particular modeling efforts and ``-ilities'' associated with them.
Particular examples of tools that rely on a metamodel to some extend include AADL \cite{mian:2015}, EST-ADL \cite{walker:2013}, and Mathworks Simulink \cite{son:2012}, to name a few.
Interoperability of these tools would require the composition of their respective implicit or explicit metamodels.

All these metamodels, including Vitech's, require the system engineer
to accept a particular paradigm \emph{and} modeling tool.
By implementing our metamodel in GraphQL we decouple the metamodel
from a particular tool and instead enforce the method
in whichever modeling language might be best equipped
to model a candidate system.
Even though the promise of metamodels is such
that they reside on top of any particular modeling abstraction, 
we note that, for example, modeling a system in SysML versus in AADL
has distinctly different objectives.
Models in SysML typically reside
at a higher level of abstraction than AADL 
but lose the expressiveness and guarantees versus modeling
in a lower-level language such as AADL.
In addition, no metamodel presented in this section
addresses safety, security, and resilience as one problem,
which is vital for modeling CPS.

\begin{figure*}[!th]
\centering
\includegraphics[width=\linewidth]{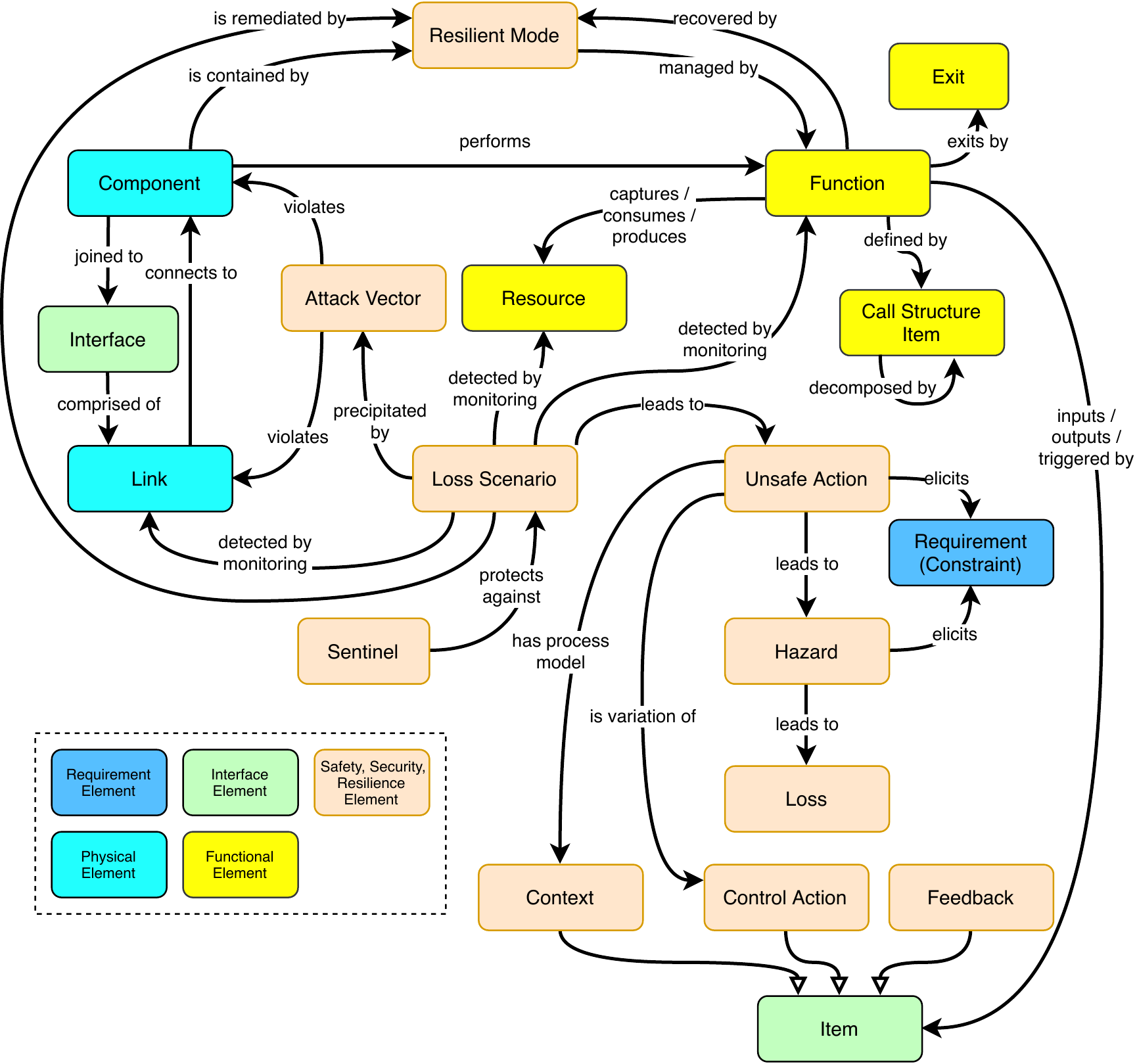}
\caption{\emph{A top-level view of the metamodel}. The proposed metamodel addresses relevant performance metrics for cyber-physical systems, namely safety, security, and resilience on top of an industry model-based systems engineering metamodel.}
\label{fig:ma-metamodel}
\end{figure*}

Our motivation was also inspired by the observation
that composing metamodels is a necessity in practice.
This led to two requirements for the metamodel.
The first is  to decouple the metamodel from a particular modeling tool
and create a flexible representation
that is designed for extensibility.
The second is to construct a compact metamodel augmentation
for a specific set of ``-ilities''
that are necessary in the design of CPS
over an already tested metamodel,
which in some sense is metamodel composition.
We believe that a number of advances have emerged
by following this design path for metamodeling.

\section{Ontological Metamodel for Safety, Security, and Resilience} 

During the early design phase of a systems lifecycle, various engineering disciplines -- hardware, software, systems, etc. -- work in parallel to refine a system design. The process is iterative in nature, with each team expressing their needs as they arise and the design morphing to incorporate the necessary modifications. The design may be documented in multiple formats to communicate and emphasize different views. Existing formats include  databases for requirements, for example IBM's Rational DOORS, diagrammatic software for architecture diagrams, text documents for interface descriptions, or model-based software for covering multiple of these representations.

The proposed metamodel (Fig.~\ref{fig:ma-metamodel}) assists both with the ease of the aforementioned design process and with maintaining the the rigor of the design. Among others, there are three anticipated benefits to using a metamodel: 
\begin{enumerate}
\item holding the various methods of documentation in alignment, which will reduce workload in maintaining the documents and prevent errors; 
\item enforcing a set structure to remain in compliance with systems engineering best practices; and 
\item connecting performance metrics like safety, security, and resilience with the system design process.
\end{enumerate}

The first benefit relates largely to viewing the metamodel outside of a specific modeling tool and applying the refinements everywhere in the design artifacts. The second benefit relates to producing an algorithmic interpretation
of an otherwise informal metamodel, such that future development of refinement types
and guarantees can be linked across modeling tool barriers.
The third benefit relates to addressing the need of assuring a number of ``-ilities'' within the metamodel, which must take a central role in the design of CPS where undesired behaviors can lead to accidents.
These three benefits summarize the difference between the Vitech MBSE metamodel
and the metamodel presented in this paper and ultimately our contributions to model federation, a significant challenge facing MBSE at large \cite{paige:2017}. 

Having said that, a MBSE metamodel should only be concerned
with the architecting activities
of system design
and not with the minutiae
of a particular system design.
It is often the case that building
on top of an existing metamodel merely leads
to added complexity both to the users 
and tool vendors \cite{fondement:2013}.
For this reason we keep the main entities of the original metamodel as-is
and construct \emph{relationships} between the main entities
of system design and the entities related
with safety, security, and resilience.
This does not mean that we enforce which parts
of the metamodel must be modeled.
Modeling is as much a form of art as it is a practice of engineering and science.
The metamodel exists to add extra structure \emph{only}
to the elements that a system designer needs to use.

\begin{table*}[!t]
\renewcommand{\arraystretch}{1.5}
\centering
\caption{The standard model-based systems engineering metamodel entities capture the fundamental components and relationships in the systems engineering lifecycle.}
\begin{tabular}{l l p{3.5in}} 
    \toprule
    Element & Entity & Description \\
    \midrule
     Requirement & \texttt{Requirement} & A \texttt{requirement} is either an originating \texttt{requirement} extracted from
    source documentation for a system, a refinement of a higher-level \texttt{requirement}, a
    derived characteristic of the system or one of its subcomponents, or a design
    decision.\\
    \midrule
    Physical &  \texttt{Component} & A \texttt{component} is an abstract term that represents the physical or 
    logical entity that performs a specific \texttt{function} or \texttt{functions}. \\
    & \texttt{Link} & A \texttt{link} is the physical implementation of an \texttt{interface}. \\ \midrule

    Interface & \texttt{Interface} & An \texttt{interface} describes the logical connections between parts of an architecture. \\
    & \texttt{Item} & An \texttt{item} represent flows within and between \texttt{functions}. An \texttt{item} is an input to or an output from a \texttt{function}. \\ \midrule
    
    Functional &  \texttt{Function} & A \texttt{function} is a transformation that accepts one or more inputs 
    (\texttt{items}) and transforms them into outputs (\texttt{items}). \\
    & \texttt{Call Structure Item} &   Recursive \texttt{call structure}, for example, select, parallel, loop, for each \texttt{function}. \\[1em]
    & \texttt{Exit} & An \texttt{exit} identifies a possible path to follow when a processing unit 
    completes. \\
    & \texttt{Resource} & A \texttt{resource} is an element, for example, power, MIPS, interceptors, that
    the system uses, captures, or generates while it is operating. \\ \midrule
    Miscellaneous & \texttt{Constraint Definition} & A \texttt{constraint definition} captures a parametric constraint as an expression, identifying the independent variable(s), with associations to the system parameters. \\ \bottomrule

\end{tabular}
\label{table:mbse-entities}
\end{table*}  

\subsection{Metamodel Description}
\label{section-ontological-metamodel}
% describe metamodel, where it came from, why is it reasonable, augmentations for Mission Aware

A key concern of any systems engineering model is an understanding of the system's architecture,  including its \texttt{components}, logical \texttt{interfaces}, and physical \texttt{links} which connect them.  Components may include hardware elements, software elements, external systems, and/or humans. Of equal concern is an understating of the expected behavior of the system being modeled. Behavior elements include \texttt{functions}, their input and output \texttt{items} as well as any \texttt{resources} provided or consumed. The \texttt{call structure} provides an understanding of behavior control flow including looping, parallel execution, path selection with \texttt{exit} choices, etc. Components \texttt{perform} functions thereby linking the physical architecture with the behavior model. These standard system modeling entities define the engineering process itself and provide structure to the essential design artifact of the system under design (Table \ref{table:mbse-entities}).

However, MBSE entities and relationships do not address ``-ilities'' necessary for the design of CPS.
Additional entities for safety, security, and resilience that are specifically related to CPS must be added
to provide evidence for the correct behavior of CPS.
Such performance metrics are defined in the augmentation of the CPS metamodel
and related to already standardized MBSE entities with properly defined relationships.
This is an important addition to the standard metamodel provided by Vitech
and defines our main contribution to metamodeling for CPS.
By adding structure to performance metrics we are better able
to design CPS that provide operational assurance
in the face of hazards or security violations.
Moreover, we provide a framework to design in and assess the efficacy of resilience,
when perimeter-based defenses are either insufficient or too costly.

\begin{table*}[!th]
\renewcommand{\arraystretch}{1.5}
\centering
\caption{The metamodel augmentations to the standard model-based systems engineering entities address safety, security, and resilience.}
\begin{tabular}{l l p{3.5in}} 
    \toprule
    Element & Entity & Description \\
    \midrule
    Control Structure & 
    \texttt{Control Action} & A controller provides \texttt{control actions} to control some process
    and to enforce constraints on the behavior of the controlled process. \\
    & \texttt{Feedback} & Process models may be updated in part by \texttt{feedback} used to observe 
    the controlled process. \\
    & \texttt{Context} & The set of process model variables and values. \\ 
    \midrule
    Hazard Analysis &  \texttt{Loss} & A \texttt{loss} involves something of value to stakeholders. Losses may include a 
    loss of human life or human injury, property damage, environmental pollution, 
    loss of mission, loss of reputation, loss or leak of sensitive information, or 
    any other loss that is unacceptable to the stakeholders. \\
    & \texttt{Hazard} & A \texttt{hazard} is a system state or set of conditions that, together with a 
    particular set of worst-case environmental conditions, will lead to a \texttt{loss}. \\
    & \texttt{Unsafe Control Action} & An \texttt{unsafe control action} is a \texttt{control action} that, in a 
    particular \texttt{context} and worst-case environment, will lead to a \texttt{hazard}. \\
    & \texttt{Loss Scenario} & A \texttt{loss scenario} describes the causal factors that can lead to the \texttt{unsafe control} and to \texttt{hazards}.    Two types of \texttt{loss scenarios} must be considered:  
    a) Why would \texttt{unsafe control actions} occur?  b) Why would \texttt{control actions} be 
    improperly executed or not executed, leading to \texttt{hazards}? \\ \midrule

    Mission Aware & \texttt{Attack Vector} & A path or means by which a hacker can gain access to a
    computer or network server in order to deliver a payload or malicious outcome.
    Attack vectors enable hackers to exploit system vulnerabilities, including the
    human element. \\
    & \texttt{Resilient Mode} & A configuration of a target system that remediates one or more 
    \texttt{loss scenarios}. \\
    & \texttt{Sentinel} & A highly secure subsystem responsible for monitoring and
    reconfiguration of \texttt{resilient modes} for a target system. \\ \bottomrule

\end{tabular}
\label{table:ma-entities}
\end{table*}

Traditional system performance metrics are captured as \texttt{parameters} of links, components, and/or functions with a \texttt{constraint definition} defining the equations and relationships between individual entities. Consideration of safety, security and resilience performance metrics require augmentation of the standard MBSE metamodel with additional concepts to capture both an operational risk perspective and an adversarial attacker perspective (Tables \ref{table:ma-entities} and \ref{table:vulnerability-assesment}).

\begin{table*}[!t]
\renewcommand{\arraystretch}{1.5}
\caption{Further refinement which captures security considerations for vulnerability assessment.}
\centering
\begin{tabular}{l l p{2.5in}} \toprule
    Entity (Association) & Attribute & Description\\ \midrule
    Attack Vector & \texttt{description} & The standard description of the attack, which often comes from databses that contain attack patterns, weaknesses, and vulnerabilities. \\ 
    & \texttt{domainOfAttack} & The domain of the attack; that is, software, hardware, communication, supply chain, social engineering, and physical security.\\
    & \texttt{outOfScope} & A binary value (true or false) flagging if the attack vector should be projected over the system.\\ 
    & \texttt{outOfScopeJustification} & A justification for not considering the attack vector within the system scope.\\ \midrule
    Attack Vector (violates) & \texttt{description} & \texttt{Component} and/or \texttt{link} specific attack vector description. \\
    & \texttt{mitigationType} & The attack vector system mitigation for the associated \texttt{component} and/or \texttt{link}, for example, defensive, diverse redundant, or hardening solution. \\
    & \texttt{justification} & A description of selected mitigation. \\ \midrule
    Loss Scenario & \texttt{detectPattern} & The \texttt{Sentinel} design pattern associated with the \texttt{Loss Scenario} (Changing Control Input, Data Consistency, Introspection).\\ 
    & \texttt{threatCategory} & The category of a threat -- terminology reused from STRIDE (Spoofing, Tampering, Repudiation, Information disclosure, Denial of Service, Elevation of Privilege)~\cite{hernan:2006}.\\ \midrule
    \makecell[l]{Loss Scenario \\(detected by monitoring)} & \texttt{constraint} & Constraint (=, \textless, \textgreater) for associated \texttt{function, link, resource} monitored by \texttt{sentinel}.\\ \bottomrule
\end{tabular}
\label{table:vulnerability-assesment}
\end{table*}

Safety and security often require specification of system behavior as a set of feedback control loops. As such, specializations of \texttt{control action} and \texttt{feedback} are provided as sub-types of the standard function input-output \texttt{item}.
While this phraseology is borrowed from STAMP, it applies to a large number of safety and security methods.
STAMP in some sense distills any general framework for ``-ilities'' at a higher abstraction level -- by leveraging notions of uncontrolled actions and control hierarchy -- that is suited for use in a metamodel. 
Specifically, \texttt{losses}, \texttt{hazards}, and \texttt{unsafe actions} are captured and related (by means of \texttt{leads to}) as part of a methodical operational risk assessment process and are recurring in a large number of safety, security, resilience, and risk management literature. Additionally, explicit associations are captured to understand an unsafe action as a \texttt{variation of} a specific control action with the \texttt{process model} system state that provides the \texttt{context} for the control action to become unsafe, which is borrowed from the domain of control theory and governs all CPS to some extent.

An important step in assessing any performance metric is to first identify \texttt{loss scenarios} which can lead to unsafe actions.
These loss scenarios are the complement of the stakeholders's requirements or otherwise define the mission of the system.
In the domain of CPS unsafe behavior and security violations are intertwinded, meaning that an attacker could transition the system to a hazardous state.
To augment the safety loss scenarios, we use databases, for example MITRE CAPEC \cite{CAPEC}, which contain \texttt{attack vectors}.
Attack vectors define the steps necessary to attack a particular system component or set of components.
By using this information from the databases it is possible to apply security information to the system architecture noting loss scenarios that could be \texttt{precipitated by} particular attack vectors. The metamodel relates the notion of loss scenario with the notion of recovery and resilience by identifying how a \texttt{sentinel} could \texttt{protect against} the loss by first indicating how it can be \texttt{detected by monitoring} a link, resource or function and then how it can be \texttt{remediated} by a specific \texttt{resilient mode}.

Augmenting the meta-model for these performance metrics, in the form of losses, attack vectors, and resilient modes has the added benefit of facilitating tradespace analysis. The metamodel does not \emph{do} the tradespace analysis per se, but instead relates the expert and operator perspectives, which are required for \texttt{priority} ranking of system losses, \texttt{likelihood} and \texttt{severity} determination for attack vectors in order to evaluate the \texttt{effectiveness} and \texttt{complexity} of resilient modes. Other import evaluation metrics for system resilient modes are an understanding of the \texttt{operational impact} and the time budget for system recovery.  Recovery time includes \texttt{detection time}, \texttt{isolation time} and \texttt{restore time} including any \texttt{operator decision time}. System simulation can evaluate the \texttt{recovery ratio} for critical system functions under various system loads and simulated attack patterns. Tradespace analysis, based on resilience metrics, enables specification of a system which responds to safety and security violations, while achieving operational priorities, within programmatic cost and time constrains (Table \ref{table:resilence-assesment}).

\begin{table*}[!ht]
\renewcommand{\arraystretch}{1.5}
\caption{Further refinement that captures metrics for assessing resilience.}
\centering
\begin{tabular}{l l p{3.0in}} \toprule
    Entity (Association) & Metric & Description \\ \midrule
    Loss & \texttt{priority} &  A ranking based on blue team (operator) determination: 1. Unacceptable \texttt{loss} and highest priority to provide \texttt{resiliency}. 2. Avoid \texttt{loss} as long as \texttt{resiliency} solution does not over-complicate operation. 3. Would like to avoid \texttt{loss}, but solution needs to be incremental. 4. Lowest priority, low-cost, simplistic solutions should be considered. \\ \midrule
    Attack Vector & \texttt{likelihood} &  A ranking based on the red team determination: the \texttt{likelihood} of the attack (high, medium, low). \\
    & \texttt{severity} & Typical \texttt{severity} of this type of attack (very high, high, medium, low, very low). \\ \midrule
    Loss Scenario & \texttt{detectionTime} & Time budget to detect \texttt{loss scenario}. Architecture impact and tradeoff for \texttt{sentinel} interfaces, for example, polling-based (system and link loading) or event-based. \\
    & \texttt{operatorDecisionTime} & Time budget to isolate the \texttt{loss scenario} via system and/or \texttt{component} tests.\\ \midrule
    Resilient Mode & \texttt{complexity} & Degree of model ``contained by'' associations. Indication of development cost (high, medium, low). \\
    & \texttt{effectiveness} & Impact on remediating High \texttt{likelihood} attacks associated with High mission \texttt{priority} (high, medium, low) \\
    & \texttt{operationalImpact} & Degree of operator training need. Degree of mission interruption (high, medium, low). \\
    & \texttt{restoreTime} & Time budget to restore system function via \texttt{resilient mode}. Architecture Impact / tradeoff for Resilient Modes: Active/Active, Active/Standby (hot, warm, cold) \\
    & \texttt{operatorDecisionTime} & Time budget for operator decision time to enable \texttt{resilient mode}. 0 implies automated resilient mode. \\ \midrule
    Function (recovered by) & \texttt{recoveryRatio} & Calculated [per \texttt{Loss Scenario}] measured/expected where if $< 1$ acceptable and if  $> 1$ not acceptable. \texttt{Recovery time} includes \texttt{detection and isolation and restoration}.  \\ \bottomrule
\end{tabular}
\label{table:resilence-assesment}
\end{table*}

\subsection{Algorithmic Implementation}
% describe the implementation, design choices for the implementation, why graphql is a reasonable choice etc. the things the implementation allows us to do, e.g.,  being agnostic to modeling language

In Myers's seminal work~\cite{myers:1990} on taxonomies of visual programming,
it is argued that visual programming languages have a number of issues in practice, some of which include:
\begin{itemize}
\item the visual representation is always significantly larger
than the text representation they replace;
\item visual languages lack formal specifications;
\item visualizations often are poor representations of the actual data; and 
\item implementations lack portability of programs.
\end{itemize}

These observations are also applicable to systems modeling languages.
Database representations are often more consise and compact versus their visual counterparts.
Implementations of modeling languages like SysML lack formal semantics.
The available diagrams often are not always the best representation
of a system models abstraction.
Finally, OMG produces standards for portable SysML models, but in practice this is not the case between vendor tools.
These drawbacks show that a tool agnostic, text-based, and database-oriented implementation for containing the relational data defining the model is useful, even when most of the modeling takes place within the visual tool. 
In addition, this opens up the system model to a number of external tools
to select model entities, process the model through other means, and report on the results
of external analysis back into the initial visual tool.

For the above reasons, we implement the formal specification of the proposed metamodel in GraphQL. We provide this formal specification to the modeling community for scientific dissemination, augmentation, and ultimately use. The full CPS schema is published online as open-source software \cite{zenodo:schema}.
GraphQL is a a schema specification language, a query language and provides query results via a JSON document~\cite{graphql}. While not initially created for capturing modeling language metamodels these attributes make it an appealing language to programmatically implement metamodels.
This is particularly the case because GraphQL is an open source standard,
which is implemented in a vendor agnostic format.
Furthermore, the schema syntax is both human readable and machine parseable.
The JSON model document provides a useful interchange format 
for various system and design modeling tools.

The top-level GraphQL specification of the CPS schema defines the metamodel at it highest level of abstraction (Listing~\ref{lst:top-level}). A single query; that is, \texttt{cpsSystemModel}, returns a complete system model including both standard MBSE entities and augmentations (Section \ref{section-ontological-metamodel}). 

\begin{lstlisting}[caption=The top level of the CPS schema captures the complete metamodel specification.,label={lst:top-level}]
schema {
  query: Query
  mutation: Mutation
}
type Query {
  cpsProjects: [Project]
  cpsSystemModel(projectId: ID!): CPSsystemModel
}
type CPSsystemModel {
  project: Project
  attackVector: [AttackVector]
  component: [Component]
  context: [Context]
  controlAction: [ControlAction]
  document: [Document]
  domainSet: [DomainSet]
  exit: [Exit]
  feedback: [Feedback]
  function: [Function]
  hazard: [Hazard]
  interface: [Interface]
  item: [Item]
  link: [Link]
  loss: [Loss]
  lossScenario: [LossScenario]
  requirement: [Requirement]
  resilientMode: [ResilientMode]
  resource: [Resource]
  unsafeAction: [UnsafeAction]
  callStructure: [CallStructure]
}
\end{lstlisting}

Each entity definition follows a pattern of:
\begin{enumerate}
    \item identity (id, name, number)
    \item attributes
    \item parameters, and
    \item relations
\end{enumerate} 
and includes a formal definition, provided in a comment block, as a mechanism to enable common understanding and usage of the modeling artefacts.

The attribute and relationship definitions are extendable and specific to the entity type.  For example, the \texttt{component} entity includes a number of useful attributes  (Listing~\ref{lst:componentattr}) and relations including  both refinement, for example, \texttt{specifiedBy}, and action sets, for example, \texttt{isViolatedBy} (Listing~\ref{lst:componentrel}).

\begin{lstlisting}[caption=The component definition follows the standard entity pattern and naming convention.,label={lst:componentattr}]
type Component {
  identity: ComponentID!
  attributes: ComponentATTR
  parameters: [Parameter]
  relations: ComponentREL
}
type ComponentID {
  id: ID!
  name: String!
  number: String!
}
type ComponentATTR {
  type: ComponentType
  inventory: [String]
  clin: String
  outOfScopeAttackAnalysis: Boolean
  outofScopeJustification: String
  mission: String
  operations: [String]
  puid: String
  purpose: String
  cost: Float
  receptions: [String]
  abbreviation: String
  title: String
  description: String
}
\end{lstlisting}

\begin{lstlisting}[caption=The component relationships show the interconnections between model types.,label={lst:componentrel}]
type ComponentREL {
  builtFrom: [BuiltFromTarget]
  builtIn: [BuiltInTarget]
  connectedTo: [ConnectedToTarget]
  documentedBy: [DocumentedByTarget]
  enablesDetectionOf: [EnablesDetectionOfTarget]
  generalizationOf: [GeneralizationOfTarget]
  isViolatedBy: [IsViolatedByTarget]
  joinedTo: [JoinedToTarget]
  kindOf: [KindOfTarget]
  performs: [PerformsTarget]
  protectsAgainst: [ProtectsAgainstTarget]
  reportedBy: [ReportedByTarget]
  simulates: [SimulatesTarget]
  specifiedBy: [SpecifiedByTarget]
}
\end{lstlisting}

The algorithmic implementation allows us to decouple the metamodel from any specific modeling tool or language. A key advantage of GraphQL over a simple JSON or XML specification is the ability to extend the GraphQL schema with additional query types as well as \emph{mutation} types to enable an application programming interface (API) for updates to a system model (Listing~\ref{lst:mutations}). The system-level mutation \texttt{cpsSystemModel()} provides a mechanism to capture the delta between variants of a system specification. These variants of functionally equivalent systems allow for a tradespace analysis between alternative approaches of addressing system safety, security and resilience.

Additionally, this API could be used to create a bidirectional link between the GraphQL implementation and a modeling tool. Developers will find an abundance of GraphQL open source libraries across multiple language bindings as well as multiple cloud hosted GraphQL services. The algorithmic implementation also provides a formal framework upon which further analysis tools can be built. For example, a graph database could be populated with entities representing nodes in the graph and entity relationships representing the edges of the graph.
One application of this transformation would be to automatically propagate security violation over the hierarchy of the model after doing model-based security assessment \cite{bakirtzis:2019}.
Another could be using standard data filtering and processing tools on the model to find particular subsystem entities,
which is a significant capability in larger system models found in industry.

\begin{lstlisting}[caption={The CPS schema supports a system-level mutation as a mechanism to capture the delta between variants of a system specification.}, label={lst:mutations}]
type Mutation {
  cpsProject(project: Project_Input): Project
  cpsSystemModel(projectId: ID!, cpsSystemModel: CPSsystemModel_Input): CPSsystemModel
}

input CPSsystemModel_Input {
  attackVector: [AttackVector_Input]
  component: [Component_Input]
  context: [Context_Input]
  controlAction: [ControlAction_Input]
  document: [Document_Input]
  domainSet: [DomainSet_Input]
  exit: [Exit_Input]
  feedback: [Feedback_Input]
  function: [Function_Input]
  hazard: [Hazard_Input]
  interface: [Interface_Input]
  item: [Item_Input]
  link: [Link_Input]
  loss: [Loss_Input]
  lossScenario: [LossScenario_Input]
  requirement: [Requirement_Input]
  resilientMode: [ResilientMode_Input]
  resource: [Resource_Input]
  unsafeAction: [UnsafeAction_Input]
}
input Component_Input {
  operation: MutationOperation!
  identity: ComponentID_Input!
  attributes: ComponentATTR_Input
  parameters: [Parameter_Input]
  relations: ComponentREL_Input
}
enum MutationOperation
{
  Create
  Update
  Delete
}
\end{lstlisting}

Finally, an important benefit of using the algorithmic GraphQL implementation
of the metamodel is that often different vendors use different modeling tools.
This means that to evaluate the diverse vendor solutions the system solicitor must acquire and be familiar
with each of the tools used in the design process of each vendor.
By using the GraphQL implementation (assuming that a bridge has been built
from each individual tool -- a one time process)
then it is possible to have a common language among modeling tools
that adhere to the same set of modeling requirements based
on the metamodel.

\section{Demonstration: Oil and Gas Pipeline}
\label{sec:demo}

% populating the model different than presenting the value of the model
% possible solution showing how the relation of these three ilities makes you think of x, y, z.
Oil and gas pipeline companies share the concerns of policymakers and others regarding the potential implications of a security violation on industry assets. There are ongoing activities to protect critical infrastructure, provide reliable energy for society, and to preserve public safety and the environment. Adversaries to this industry activity include nation states, criminal organizations, and unaffiliated bad actors seeking to steal intellectual property, compromise industrial control systems, and other nefarious goals. The industry has seen the evolution of cyber criminals and the advancement of the techniques, tactics, and procedures they use, moving from manual operations to more sophisticated and widespread machine-to-machine automated attacks with use of augmented intelligence. There are multiple other attack vectors including insider threats and attacks via supply chain tampering or disruption. We show an application of the metamodel. Namely, how the metamodel allows us to capture  the mission-oriented information, the potential losses to the stakeholders, and the implications of security violations through this CPS example.

\begin{figure*}[!th]
    \centering
    \includegraphics[width=.95\textwidth, frame]{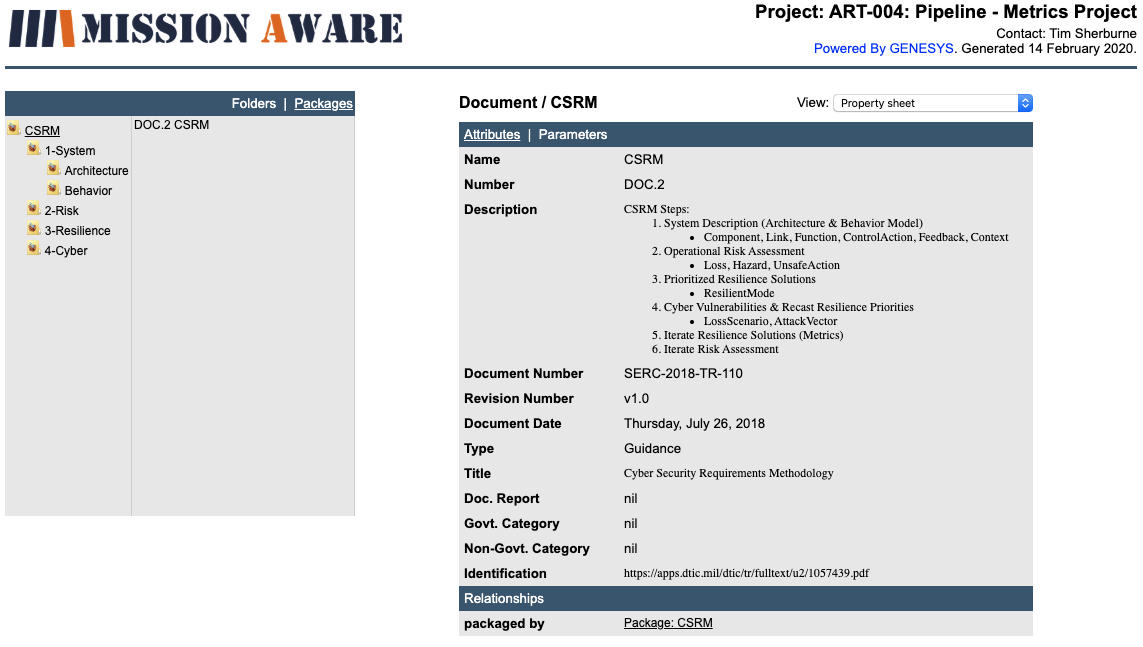}
    \caption{The top-level web-view of oil and gas pipeline demonstration maps to the top-level concepts on metamodel.}
    \label{fig:cs-webview}
\end{figure*}

This demonstration system is decomposed and organized according to the mission aware methodology using the Vitech GENESYS MBSE modeling tool which was extended with our metamodel. The particular tool is not necessary to use the metamodel but we use the tool and its associated diagrams to visualize the different model views as defined by the metamodel. Additionally, we were able to show that our metamodel extension was straight forward and that an organizational investment in a particular modeling tool can be preserved without significant tooling or retraining costs. GENESYS provides a mechanism to export a system model to a web-based team view. In this section, we present a summary of key model artifacts but we offer the complete model as open-source, which does not require to use the tool. The full oil pipeline model is available online \cite{zenodo:model}. The web-view model navigator (Fig.~\ref{fig:cs-webview}) shows a \texttt{package} view to organize the model artifacts presented in the following sections. Expanding a package folder presents a hierarchy of related entity types.
Based on the metamodel, the general overview of the system model takes the form of a system description, which contains an architecture and an associated behavior, operational risk to the system, resilience design patterns based on the operational risk, and potential threats to the system. In the following sections, we will navigate the model at each of those levels to show how the metamodel assists with adding structure to the modeling activity and how the metamodel relates these different but important views. This means that we will not focus on an exhaustive presentation of the model itself but rather show the relationship between the model and the metamodel and how it could be used to analyze and extend the system's safety, security, and resilience considerations.

As a notational choice, we denote model-specific entities with \textsf{sans serif} text font and metamodel entities with \texttt{typewriter} text font.

\begin{figure}[!p]
\centering
\begin{subfigure}{.6\textwidth}
  \centering
  \includegraphics[width=1\linewidth]{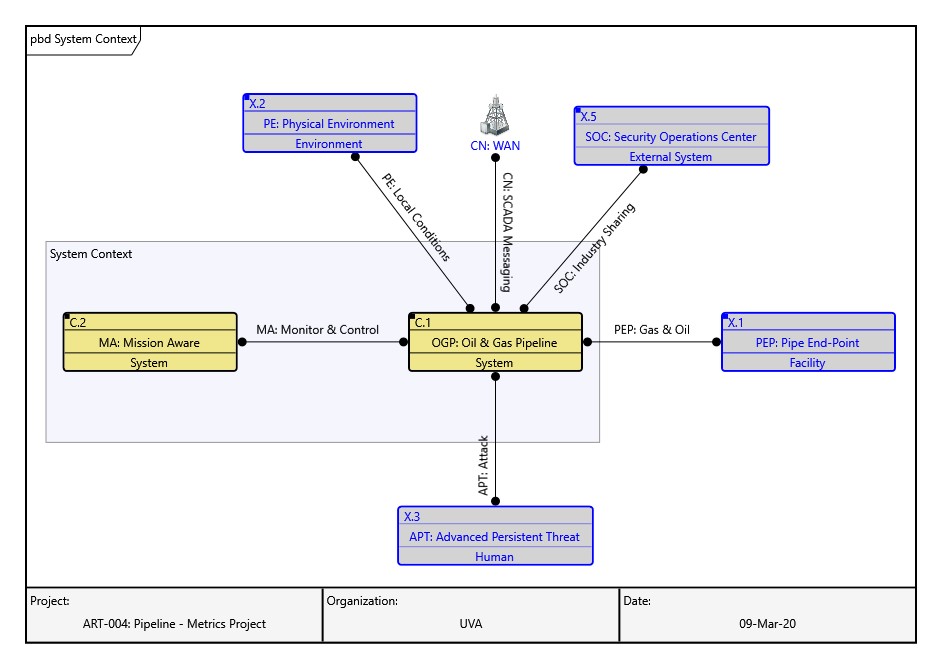}  
  \caption{The \emph{system context} physical block diagram models the boundary of the system and its external interfaces.}
  \label{fig:cs-context-pb}
\end{subfigure}
\begin{subfigure}{.6\textwidth}
  \centering
  \includegraphics[width=1\linewidth]{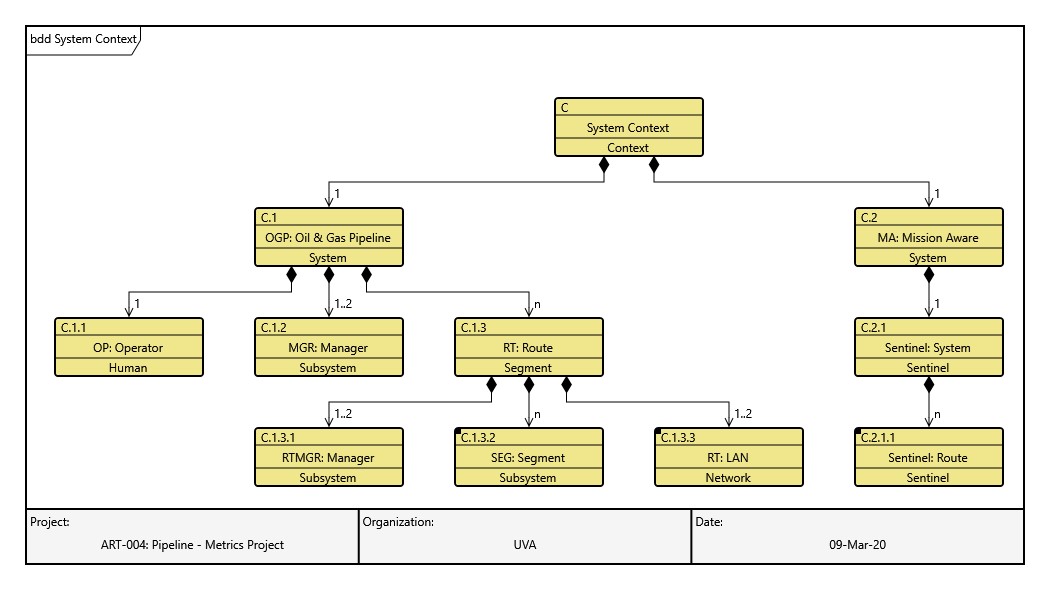}  
  \caption{The \emph{system context} block definition diagram models the decomposition of the system including cardinality of sub-components.}
  \label{fig:cs-context-bdd}
\end{subfigure}
\begin{subfigure}{.6\textwidth}
  \centering
  \includegraphics[width=1\linewidth]{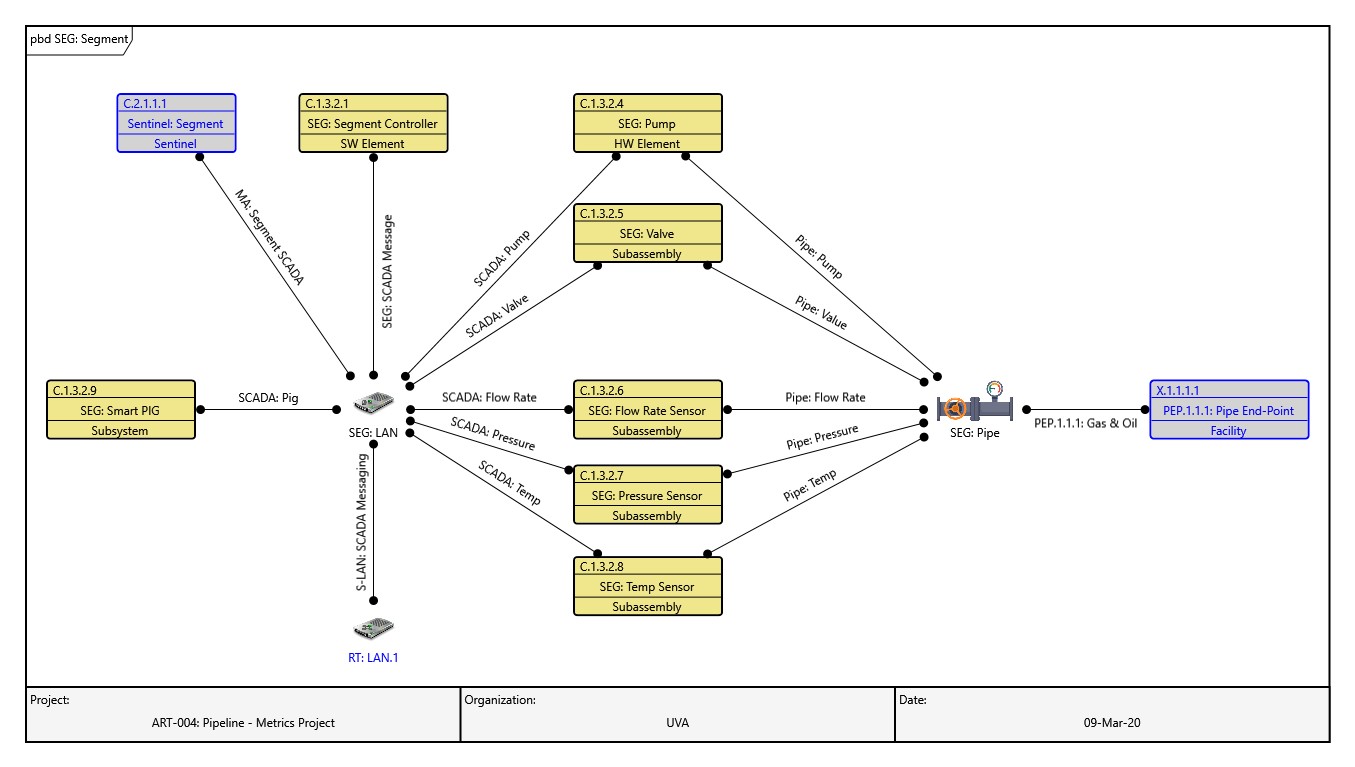}  
  \caption{The \emph{segment} physical block diagram models the physical pipe, pump, value, sensors and controller and connectivity of the SCADA LAN.}
  \label{fig:cs-segment-pb}
\end{subfigure}
\caption{The architecture of the system is hierarchically decomposed using the metamodel \texttt{component} and \texttt{link} entities and is visualized with physical block diagrams and block definition diagrams.}
\label{fig:architecture}
\end{figure}

\subsection{System Description} \label{sec:demo-system}

The system description defines the base model for the system under examination.
It is fundamentally the model in which ``-ilities'' act on or are otherwise contained in.
Therefore, it associates strongly with the MBSE entities of the metamodel (Table \ref{table:mbse-entities}).
Specifically, artifacts of the system description include the system context, the architecture of the system, and its functional behavior.

The system context physical block diagram (Fig. \ref{fig:cs-context-pb}) defines the boundaries and external interfaces for the oil and gas pipeline being evaluated. Each node on the diagram is an instance of a \texttt{component} while each connecting line is an instance of a \texttt{link} associated by the \texttt{connects to} relation. The system context diagram enables a common understanding among stakeholders of the scope of the system model. In our demonstration system the \textsf{pipeline end-points} (drilling rigs, refineries, etc.), \textsf{wide-area communications network} (cellular, satellite, etc.), \textsf{environment} (weather conditions, geography, etc.), and \textsf{security operations center}, for sharing of industry related events, are all external to our demonstration system model.  Additionally, to enable simulation of cyber attacks, an \textsf{advanced persistent threat} interface is included. Within the system context is the pipeline itself and a peer \textsf{mission aware} system (sentinel) which is concurrently evaluated and is responsible for resilient mode reconfiguration based on detected illogical system behavior which may indicate safety and/or security vulnerabilities.

The system context block definition diagram (Fig. \ref{fig:cs-context-bdd}) shows three levels of the pipeline system decomposition. Each node on the diagram is an instance of a \texttt{component} while each connecting line is a \texttt{built from} association between components, showing the cardinality of each sub-component. In our demonstration system, the oil and gas pipeline contains a single \textsf{human operator}, one or two (if redundant) \textsf{system managers} and $n$ \textsf{pipeline routes}. In turn, a pipeline route contains one or two (if redundant) \textsf{route managers}, one or two (if redundant) \textsf{route LANs}, and $n$ \textsf{pipeline segments}.

Finally, a physical block diagram (Fig. \ref{fig:cs-segment-pb}) defines the architecture of a pipeline \textsf{segment}.  Each segment includes the physical \textsf{pipe}, \textsf{pump}, and \textsf{valve} that deliver oil and gas from one endpoint to another. A set of sensors (\textsf{pressure}, \textsf{temperature}, \textsf{flow rate}) provide feedback to the \textsf{segment controller} on the operational state of the segment.  A \textsf{segment LAN} connects segment components for SCADA messaging and also provides connectivity to the higher-level \textsf{route LAN}. As part of the regular maintenance process for a pipeline segment (known as ``pigging'' within the industry), a \textsf{smart PIG} is periodically used to clean and inspect the pipe.

\begin{figure}[!t]
\centering
\begin{subfigure}{.49\textwidth}
  \centering
  \includegraphics[width=1\linewidth]{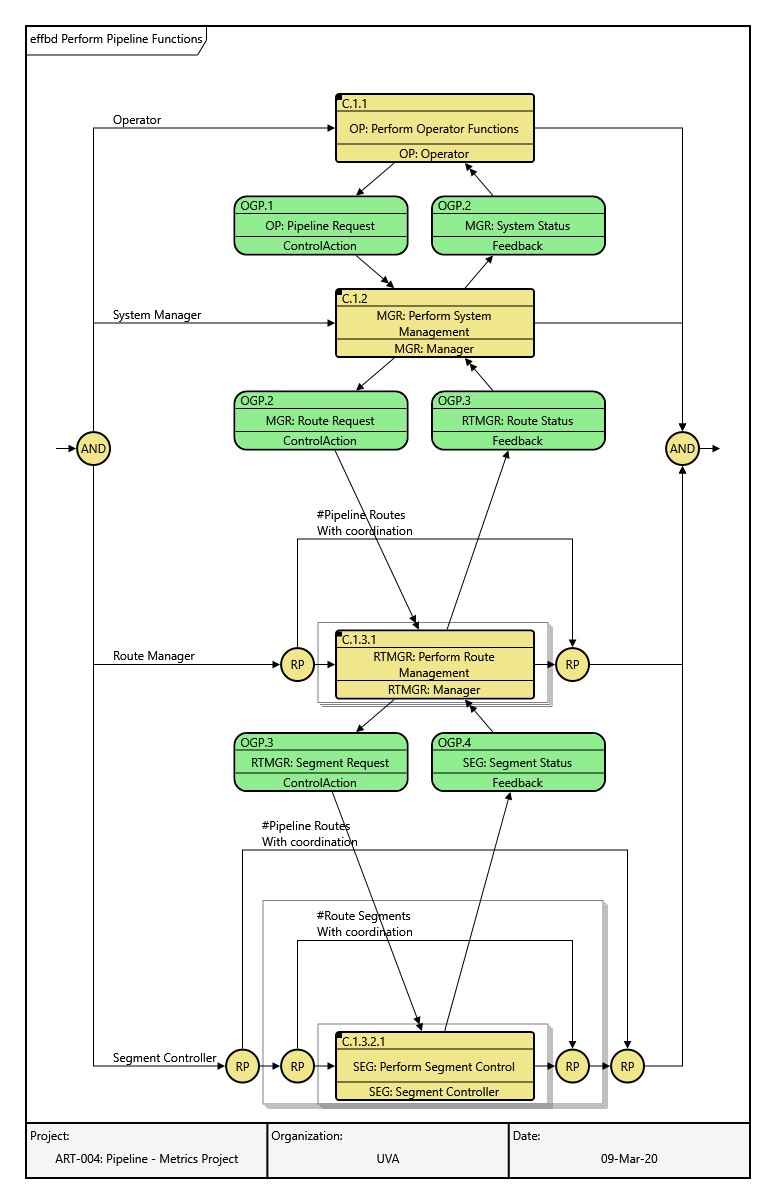}  
  \caption{The \textsf{perform pipeline function} EFFBD models the hierarchical feedback control structure of the oil and gas pipeline.}
  \label{fig:cs-pipeline-efbdd}
\end{subfigure}
\begin{subfigure}{.49\textwidth}
  \centering
  \includegraphics[width=1\linewidth]{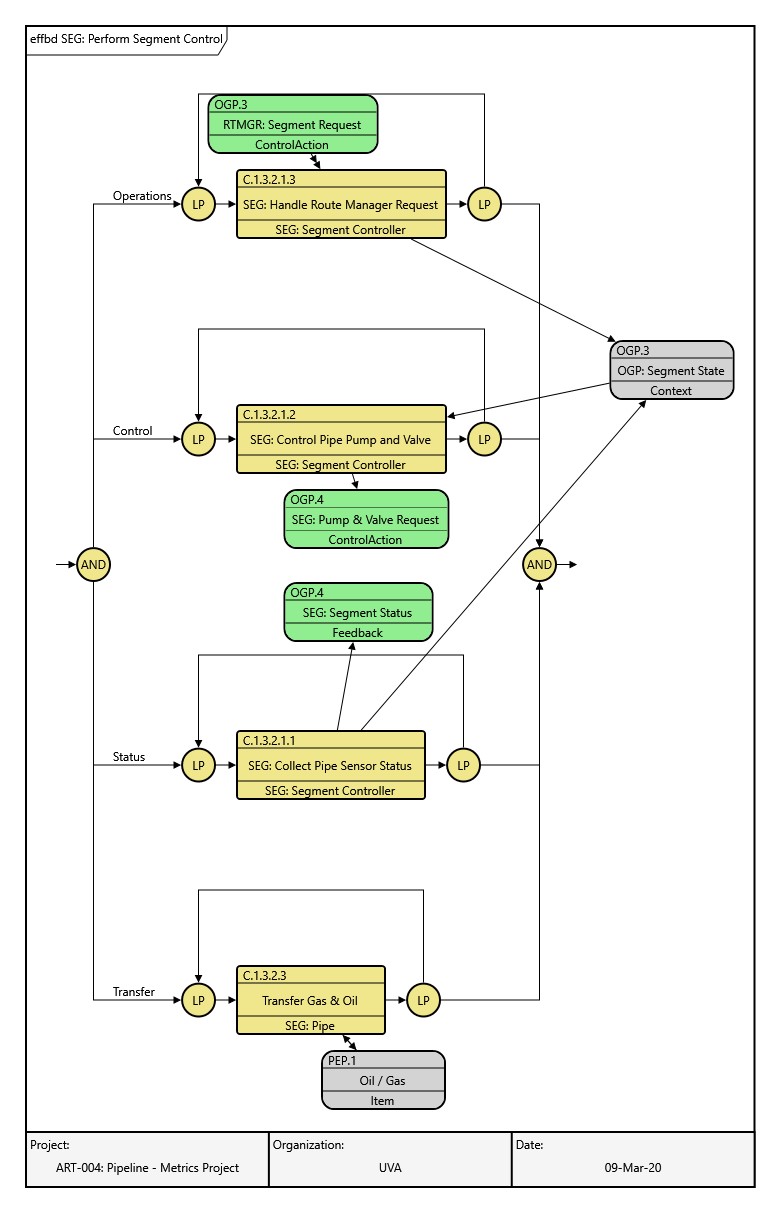}  
  \caption{The \textsf{perform segment control} EFFBD models the management of segment state to affect oil and gas transfer.}
  \label{fig:cs-segment-efbdd}
\end{subfigure}
\caption{The behavior of the system is hierarchically decomposed using the metamodel \texttt{function}, \texttt{control action}, \texttt{feedback}, \texttt{context} and \texttt{call structure} entities and is visualized with enhanced function flow block diagrams (EFFBD).}
\label{fig:behavior}
\end{figure}

Turning to system behavior, an enhanced functional flow block diagram (EFFBD) (Fig. \ref{fig:cs-pipeline-efbdd}) defines the top-level behavior for the oil and gas pipeline in the form of a feedback control structure.  The diagram shows the top-to-bottom hierarchy of the control structure. The outer-level \texttt{and} block shows that each of the \textsf{lanes} of behavior (operator, system manager, route manager, segment controller) execute in parallel. The behavior control flow logic is captured as instances of the metamodel \texttt{call structure item}. Each of the yellow rectangles represent an instance of a metamodel \texttt{function} showing the function number, name and \texttt{performed by} associated \texttt{component}. The green rounded rectangles represent instances of a metamodel \texttt{control action} or \texttt{feedback} item. The lines between functions and control actions or feedback items are represented as \texttt{outputs} or \texttt{triggers} associations. The \texttt{replicate} blocks for the Route Manager and Segment Controller indicate that there $n$ concurrent instances of these behavior blocks and are captured as call structure items that \texttt{decompose} their respective branches of the behavior model.

The \textsf{perform segment control} function is further \texttt{decomposed by} a second level EFFBD (Fig. \ref{fig:cs-segment-efbdd}) with lanes for operations, control, status, and transfer. Each of these lanes operates in parallel in a continuous loop.  The \textsf{handle route manager request} receives \texttt{control actions} from the route manager and maintains the requested state within the \textsf{segment status: context}. Based on segment state context (requested state and senor status), the \textsf{control pipe pump and value} function initiates \texttt{control actions} to the segment pump and value. The \textsf{collect pipeline sensor status} function monitors and maintains the sensor status within the \textsf{segment state: context} and then forwards that state via \textsf{segment status: feedback} to the route manager. Finally, the \textsf{transfer gas and oil} function provides the physical movement of oil and gas through the pipe segment.  

\begin{figure*}[!t]
    \centering
    \includegraphics[width=.75\textwidth]{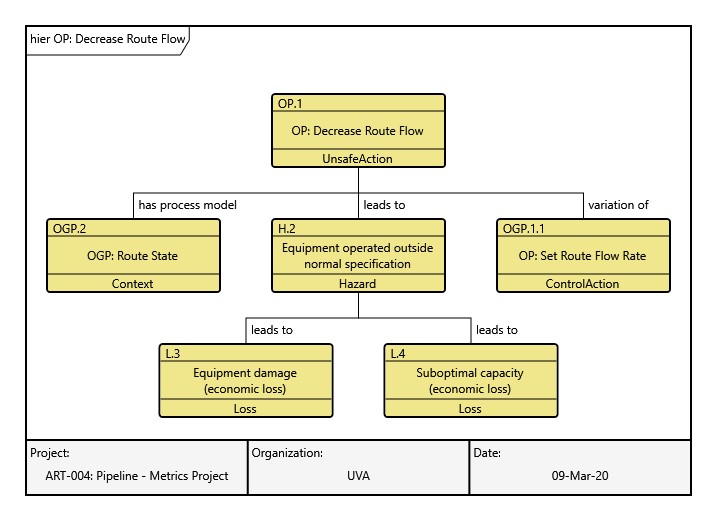}
    \caption{The operational risk of the system is described using the metamodel \texttt{loss}, \texttt{hazard}, \texttt{unsafe action} entities with associations to the system description \texttt{context} and \texttt{control action} entities and is visualized in a hierarchy diagram.}
    \label{fig:cs-decrease-flow-uca-hier}
\end{figure*}

\subsection{Operational Risk Assessment} \label{sec:demo-risk}
After the system is described to an appropriate level of detail, attention is focused on understanding operational risks for the system from the perspective of end users. Through STPA \cite{leveson2018stpa} the metamodel provides a top-down process to aid in the identification of the model artifacts for the operational risk assessment including \texttt{losses}, \texttt{hazards}, and \texttt{unsafe actions} (Table \ref{table:ma-entities}). 

The system behavior model provides an inventory of \texttt{control actions} that are methodically considered to identify potential unsafe actions.  From the pipeline behavior model, control actions include \textsf{Operator: Pipeline Request}, \textsf{Manager: Route Request}, \textsf{Route Manager: Segment Request}, and \textsf{Segment: Pump \& Valve Request}. As defined by STPA, there are four ways a control action can be unsafe: 
\begin{itemize}
  \item Not providing the control action leads to a hazard.
  \item Providing the control action leads to a hazard.
  \item Providing a potentially safe control action but too early, too late, or in the wrong order.
  \item The control action lasts too long or is stopped too soon (for continuous control actions).
\end{itemize}

As \texttt{unsafe actions} are considered and identified, the \texttt{is variation of} association is updated between the \texttt{unsafe action} and target \texttt{control action}.  This association can be used by model reporting tools to ensure that all system \texttt{control actions} have been considered and that a complete set of \texttt{unsafe actions} have been defined, especially as the system behavior model is iterated and evolved over its lifetime. Additionally, each \texttt{unsafe action} is associated via the \texttt{has process model} to a \texttt{context}, identifying the specific values that will cause the \texttt{unsafe action} to \texttt{lead to} a \texttt{hazard}. 

One example of such risk is \textsf{decrease route flow} (Fig. \ref{fig:cs-decrease-flow-uca-hier}) where flow rate for a pipeline route is decreased before achieving optimal flow as defined by the associated \texttt{process model} context \textsf{route state}. This unsafe action is a \texttt{variation of} the \textsf{set route flow rate} \texttt{control action} which can \texttt{lead to} \textsf{equipment operated outside normal specification} \texttt{hazard} which in turn can \texttt{lead to} either \textsf{equipment damage} or \textsf{sub-optimal capacity} \texttt{losses}. As part of the tradespace analysis, it is required that system operators \emph{prioritize} losses (Table \ref{table:resilence-assesment}).

Failure to fully identify \texttt{unsafe actions} can lead to an incomplete specification of resilience solutions and/or an incomplete specification of loss scenarios which are considered in the next sections.  The metamodel \texttt{is variation of} and \texttt{has process model} associations provide a mechanism to ensure a robust operational risk assessment.

Therefore, in this view of CPS modeling (through the metamodel presented in this paper) losses are the main criterion
for operational risk.
This is in alignment with much of the current literature in safety
but might strike security analysts as odd.
We posit that this is one of the benefits of the metamodel,
namely that safety, security, or resilience extension
or modification are based on the potential loss that they mitigate from,
thereby, providing evidence for their necessity and cost
to decision makers.
Further, by prioritizing losses it is possible
to compare and contrast what safety, security, or resilience considerations
should be added to the system first.

\subsection{Resilience Solutions} \label{sec:demo-resilience}
Following the operational risk assessment, system resilience solutions are proposed by system design experts. These solutions are focused on segments of the system that are within a feedback control path for related \texttt{unsafe actions} that \texttt{lead to} the highest priority system \texttt{losses}. Summarizing the feedback control loop; a \texttt{link} \textit{\texttt{connects to}} a \texttt{component}, a \texttt{component} \textit{\texttt{performs}} a \texttt{function}, a \texttt{function} \textit{\texttt{outputs}} or \textit{\texttt{is triggered by}} a \texttt{control action} or by \texttt{feedback}, and \texttt{control actions} and \texttt{feedback} are \textit{\texttt{transferred by}} a \texttt{link}. Resilient solutions are prioritized for \texttt{components} and \texttt{links} that address the highest quantity of, and most important, \texttt{unsafe actions}. The \textsf{decrease route flow} unsafe action is a \texttt{variation of} the \textsf{set route flow rate} \texttt{control action}. From the behavior model, it is shown that the feedback control loop for this \texttt{control action} includes the \textsf{Operator}, \textsf{Manager}, \textsf{Route Manager}, \textsf{Segment Controller}, \textsf{Pump \& Valve}, and the \textsf{Sensor}  \texttt{components}.  From the architecture model, it is shown that these \texttt{components} are connected by the \textsf{WAN: SCADA Messaging}, \textsf{Route: SCADA Messaging}, and the \textsf{Segment: SCADA Messaging} \texttt{links}. Resilience solutions are considered for all of these \texttt{components} and \texttt{links}. Without these metamodel associations, important resilient modes could be missed or the value of proposed resilient modes may not be appreciated during the tradespace analysis leading to a less resilient system which in turn may cause \texttt{hazardous} system states \texttt{leading to} unacceptable system \texttt{losses}

\begin{figure*}[!th]
    \centering
    \includegraphics[width=\textwidth]{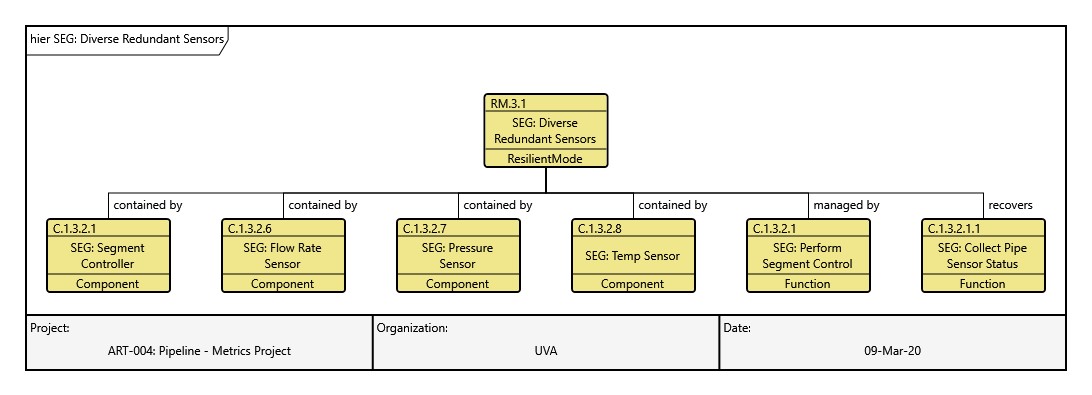}
    \caption{Resilience solutions for the system are described using the metamodel \texttt{resilient mode} with associations to the system description \texttt{component} and \texttt{function} entities and is visualized in a hierarchy diagram.}
    \label{fig:cs-sensor-redundancy-heir}
\end{figure*}

Design patterns for resilience solutions include diverse redundancy (which limits the effectiveness of insider or supply chain attacks), hardened design, perimeter defense, etc. An example of a resilience solution is \textsf{diverse redundant sensors} (Fig. \ref{fig:cs-sensor-redundancy-heir}) where the \textsf{segment controller} and \textsf{sensors} are \texttt{contained by} the solution. The degree of contained by associations are an indication of implementation complexity (Table \ref{table:resilence-assesment}) and provide an additional aid in tradespace analysis. The solution \texttt{recovers} the \textsf{collect pipe sensor status} \texttt{function} and is \texttt{managed by} (enable or disable) the \textsf{perform segment control} \texttt{function}. The control function could be performed by an operator or, if desired, automated by a sentinel.

\subsection{Vulnerability Assessment} \label{sec:demo-vulnerability}
Guided by safety and security experts a system vulnerability assessment is performed next. Identification of \texttt{loss scenarios} is the primary metamodel artifact that captures this assessment. STPA provides a structured approach for identifying loss scenarios by analysis of the system feedback control structure while security experts will also consult databases of \texttt{attack vectors} considering how a \texttt{loss scenarios} could be \texttt{precipitated by} these attack vectors. The loss scenario is linked to the operational risk assessment using the \texttt{leads to: unsafe action} association and to a resilience solution using the \texttt{is remediated by: resilient mode} association. To enable a \texttt{sentinel} monitor, a \texttt{detected by monitoring} association to a system \texttt{link, resource or function} is defined. The inventory of \texttt{unsafe actions} and associated \texttt{context} that lead to \texttt{hazardous} system states provide valuable insight into illogical system behavior, that a \texttt{sentinel} could detect and report, as an indication of safety and/or security vulnerabilities. Identification of these monitoring needs early in the system design life cycle ensures that appropriate interfaces are incorporated in the system design and are not an after thought. For example, as defined by the \textsf{decrease route flow} \texttt{unsafe action}, this \texttt{control action} is considered unsafe if received when the \texttt{process model} \textsf{currentFlowRate} is less than 80\% of \textsf{targetFlowRate} (i.e. control action received too soon). One cause of this \texttt{unsafe action} could be falsified sensor \texttt{feedback}.  For a sentinel to detect this situation, it would require interfaces to both the \textsf{targetFlowRate} and to real-time senor reports. The sentinel \texttt{data consistency} design pattern looks for inconsistencies between values as an indicator of safety and / or security vulnerabilities. 

\begin{figure*}[!th]
    \centering
    \includegraphics[width=\textwidth]{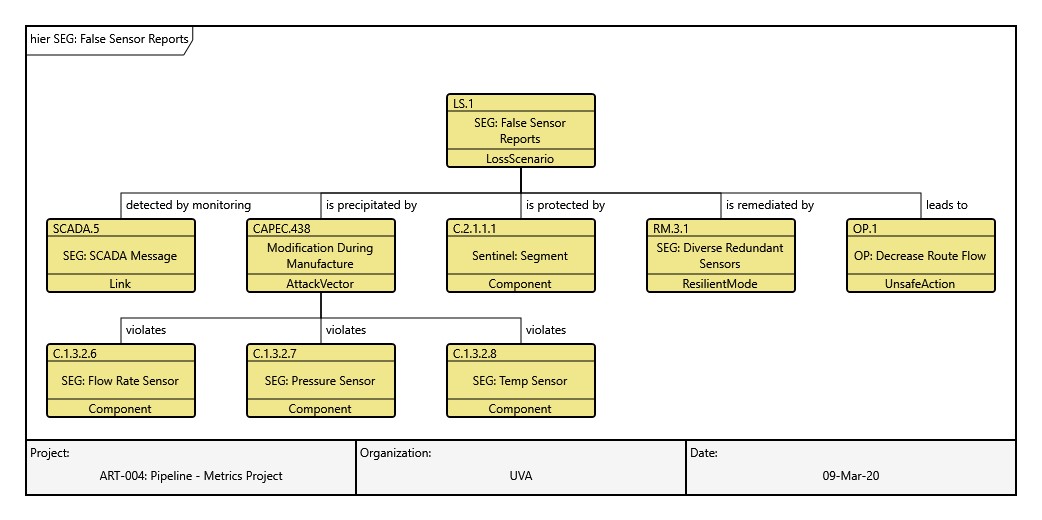}
    \caption{The vulnerability assessment for the system is described using the metamodel \texttt{loss scenario and attack vector} with associations to the operational risk assessment (\texttt{leads to}) and resilience solutions (\texttt{is remediated by}) and is visualized in a hierarchy diagram.}
    \label{fig:cs-false-sensor-loss-scenario-hier}
\end{figure*}

An example of a loss scenario is \textsf{false sensor reports} (Fig. \ref{fig:cs-false-sensor-loss-scenario-hier}) where the loss scenario is \texttt{detected by monitoring} \textsf{SCADA message: link}, \texttt{is precipitated by} \textsf{modification during manufacture: attack vector}, \texttt{is remediated by} \textsf{diverse redundant sensors: resilient mode}, and can \texttt{lead to} \textsf{decrease route flow: unsafe action}. The analysis leading to this loss scenario specification are largely the same from both the safety and security perspectives.  It is the consideration of intentional actions (\texttt{attack vectors}) that differentiates the security analysis. The benefit of this overlap ensures that the identified resilient mode and detection mechanism adequately address both safety and security concerns in the most efficient way. The vulnerability assessment is also responsible for determining appropriate values for the \texttt{likelihood and severity} of \texttt{attack vectors} as well as time budgets for \texttt{detecting}, \texttt{isolating} and \texttt{restoring} the system (Table~\ref{table:resilence-assesment}) which will be leveraged in the following iterative tradespace analysis.

\subsection{Iterative Tradespace Analysis}
Systems engineering is an inherently iterative process and balancing the perspectives of operational risk and system vulnerability also require iteration to achieve an optimal system solution within programmatic budget and time constraints. The resilience metrics (Table \ref{table:resilence-assesment}) provide a framework for evaluating the effectiveness of resilience solutions in response to safety and security violations while achieving operational priorities. Determining an appropriate resilience solution for a critical subsystem will likely have multiple approaches, for example, deciding to design redundancy or add security hardening, with security experts preferring one approach while system operators possibly preferring another due to usability considerations.  The metamodel provides a mechanism for all stakeholders to understand the trade-offs and a place to document agreements and the process used to reach consensus.  This documentation is invaluable to future system enhancements, evolution and maintenance which likely involves different team members.  As another example, a system architecture supporting a diverse redundant subsystem must assure that the system recovery time budgets are met. A polling-based detection mechanism may prove to be insufficient and may instead require an event-based notification solution to achieve the required detection time under various system loads.

\section{Conclusion}
In this paper, we extend an industry  metamodel
to address the safety, security, and resilience of CPS
in a unified framework.
The promise of a unified metamodel for all system design has proven
to assist little in the assurance of different ``-ilities''.
This is partially because different types of systems prioritize ``-ilities''
based on their operational needs.
For the domain of CPS it is vital that safety, security,
and resilience are considered during the design phase
of the lifecycle, where violation of these three ``-ilities''
can lead to accidents.

A unification
of three technologies was needed to design
the metamodel: (1) a concrete approach to MBSE,
(2) a safety and security method that is grounded
on system losses, and (3) a structured approach
to mitigation.
Additionally, we found that restricting the metamodel
to a particular language or tool is insufficient
given the diverse sources of generating model artifacts
at the design phase.
For this reason we algorithmically implement our metamodel
in GraphQL such that it is agnostic to a particular modeling language or tool.

By implementing this unification in a concrete metamodel
we facilitate consistency between CPS model views;
coordination of safety, security, and resilience
with system models;
and tradespace analysis of these three metrics
in relation to candidate design solutions.
We demonstrate these results and how they are modeled
in a pragmatic setting in a demonstration
of an oil and gas pipeline.
The GraphQL implementation of the metamodel
has also allowed us to interface
with other analysis methods,
such as model-based security assessment
outside of the particular modeling tool we use.
This capability can be extended
to populate other tools and methods
based on one model, the results of which can then be
reimported to a single modeling tool.
Therefore, achieving bidirectionality between modeling and different types 
of analysis -- 
a missing capability currently for a vast majority of tools and methods.

\section{Acknowledgments}

This material is based, in part, upon work supported by the Stevens Institute of Technology through SERC under USDOD Contract HQ0034-13-D-0004. SERC is a federally funded University affiliated research center managed by Stevens Institute of Technology. Any opinions, findings and conclusions or recommendations expressed in this material are those of the authors and do not necessarily reflect the views of the USDOD.

\bibliographystyle{plainnat}
\bibliography{manuscript}

\end{document}